\documentclass{vgtc}                          
\pdfoutput=1

\graphicspath{{figures/}{pictures/}{images/}{./}} 

\usepackage{times}                     

\usepackage{tabu}                      
\usepackage{booktabs}                  
\usepackage{lipsum}                    
\usepackage{mwe}                       

\usepackage{amsmath}
\usepackage{amssymb}
\usepackage{amsfonts}
\usepackage{enumitem,kantlipsum}
\DeclareMathOperator*{\argmax}{arg\,max}

\usepackage{wrapfig}

\usepackage{textcomp}
\usepackage{scalerel}

\usepackage{mathptmx}                  

\onlineid{0}

\vgtccategory{Research}

\vgtcinsertpkg

\title{The Visualization JUDGE : Can Multimodal Foundation Models Guide Visualization Design Through Visual Perception?}

\author{Matthew Berger\thanks{e-mail: matthew.berger@vanderbilt.edu}\\ %
        \scriptsize Vanderbilt University %
\and Shusen Liu\thanks{e-mail: shusen.liu.hust@gmail.com}\\ %
     \scriptsize Lawrence Livermore National Laboratory}
     
\abstract{
Foundation models for vision and language are the basis of AI applications across numerous sectors of society.
The success of these models stems from their ability to mimic human capabilities, namely visual perception in vision models, and analytical reasoning in large language models.
As visual perception and analysis are fundamental to data visualization, in this position paper we ask: how can we harness foundation models to advance progress in visualization design?
Specifically, how can multimodal foundation models (MFMs) guide visualization design through visual perception? 
We approach these questions by investigating the effectiveness of MFMs for perceiving visualization, and formalizing the overall visualization design and optimization space.
Specifically, we think that MFMs can best be viewed as judges, equipped with the ability to criticize visualizations, and provide us with actions on how to improve a visualization.
We provide a deeper characterization for text-to-image generative models, and multi-modal large language models, organized by what these models provide as output, and how to utilize the output for guiding design decisions. 
We hope that our perspective can inspire researchers in visualization on how to approach MFMs for visualization design.
} 


\begin{document}


\firstsection{Introduction}

\maketitle

\begin{figure}[!t]
    \centering
    \includegraphics[width=1\linewidth]{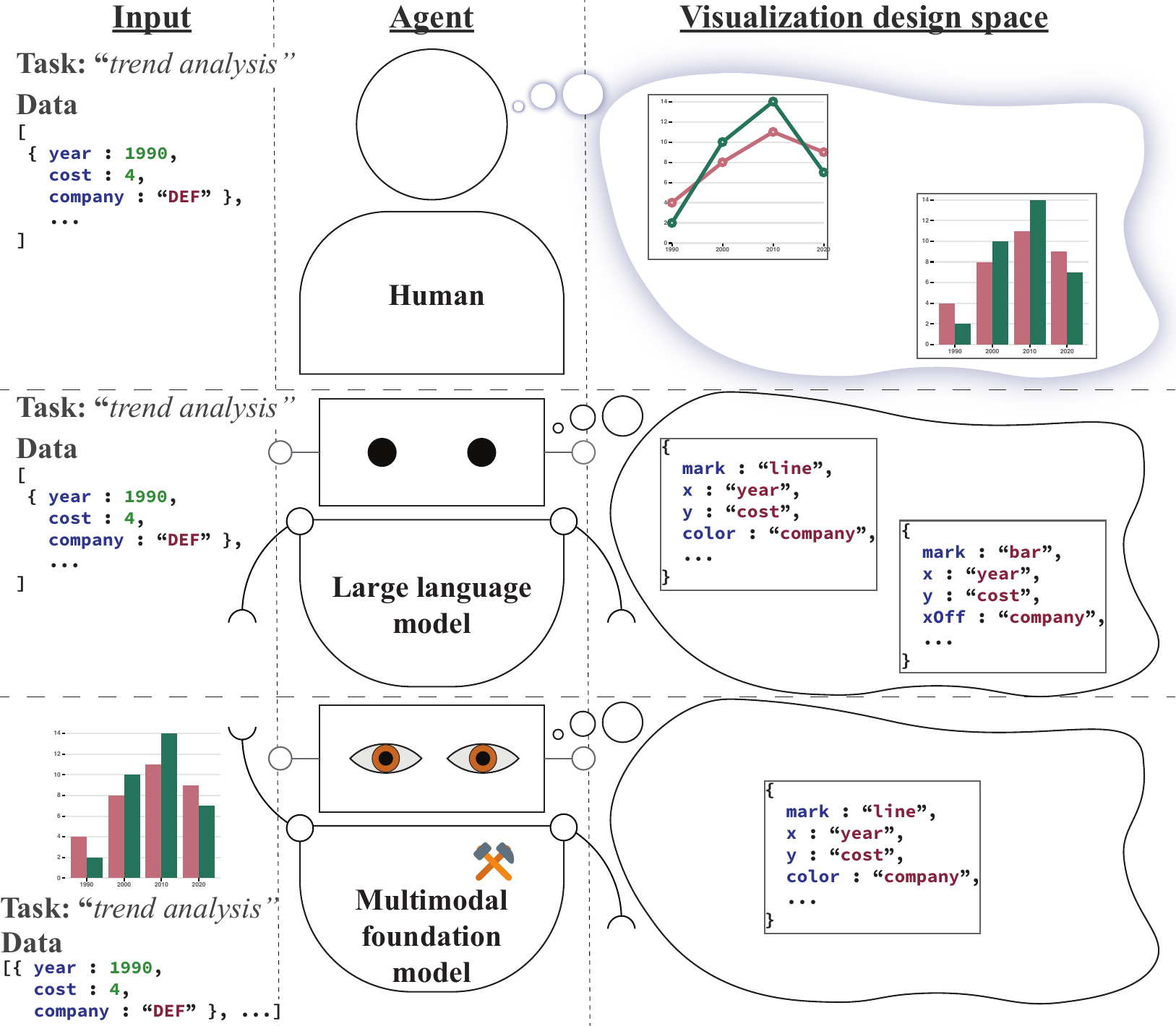}
    \caption{Unlike large language models (middle), multimodal foundation models (MFM) (bottom) can process both language and vision. Where humans can think about possible visual encodings for a particular dataset \& task (top), MFMs can similarly reason about visualizations, represented as images, in designing visualizations.}
    \label{fig:human-llm-mllm}
\end{figure}

Visualization design is all about choices.
The choices we make about how to represent data, choices on the visual encoding of data, the spatial arrangement of views, and designed interactions, amongst many other factors.
Visualization design is thus often viewed as \emph{search}~\cite{sedlmair2012design}, in which our design choices provide us with possible actions to take.
More specifically, the search can be viewed as looking for a visualization, over a large space of possible designs, that satisfies a set of user goals (c.f. Fig.~\ref{fig:human-llm-mllm}-top).
Often these goals are centered on analytical tasks that concern our data, though other factors such as aesthetics \cite{zangwill2003aesthetic} and memorability \cite{borkin2013makes,rust2020understanding} can also play a role.

Ideally, each choice we make as part of this search process brings us closer to our goals.
But how do we know we are actually making progress?
This typically requires us to (\textbf{A1}) perceive our current visualization, (\textbf{A2}), reflect on the visualization's strengths and weaknesses, and (\textbf{A3}) update our design choices to improve on the identified weaknesses.
Visual perception (\textbf{A1}, \textbf{A2}) is essential to this process: we know a good visualization when we see it.
Yet what is challenging, and often time-consuming, about design is how to make choices that lead us to good visualizations (\textbf{A3})~\cite{bigelow2014reflections}.

Given the rise in generative AI, it is natural to consider AI models that are capable of generating data visualizations as a means of exploring our options (\textbf{A3}).
To this end, a flurry of recent work~\cite{zhang2023adavis, chen2023beyond, ko2024natural, yang2024matplotagent, tian2024chartgpt, luo2021natural, maddigan2023chat2vis, zhao2024leva} has investigated the use of large language models (LLMs) for generating visualizations.
For applications of visualization design, the output of an LLM is text, typically a visualization specification~\cite{luo2021natural} or executable code~\cite{yang2024matplotagent} (c.f. Fig.~\ref{fig:human-llm-mllm}-middle).
Moreover, the input is text, often limited to some characterization of the data, for instance, a list of dataset attributes, a summary of the data distribution for each attribute, or simply raw tabular data.
However, the ability of an LLM to reason about a visualization is limited by the text given as input.
In order to make well-informed critiques on a visualization (\textbf{A2}), it is crucial to \emph{perceive the image} produced by a visualization (\textbf{A1}).
The visual patterns that can be rapidly detected in a visualization, and used to draw conclusions about our data, are just as important as the specifications used to define a visualization.

The aim of this paper is to characterize the use of multimodal foundation models (MFM), namely models that can reason about language and vision, for guiding visualization design.
Just as visual perception is essential to humans in making design choices (c.f. Fig.~\ref{fig:human-llm-mllm}-top), we believe that the vision capabilities of MFMs (c.f. Fig.~\ref{fig:human-llm-mllm}-bottom) offer a promising, yet under-explored, means of visualization reasoning.
Machine perception gives the necessary piece for AI models to approach design in a similar way to humans: (\textbf{A1}) a model can process, as input, an image of a visualization, (\textbf{A2}) judge the visualization based on the specified goals of the user, and (\textbf{A3}) provide some kind of signal about how to improve the visualization.
Similar to how humans must \emph{search} over a space of visualizations, MFMs give the opportunity to frame design as \emph{optimization}.
And although precedence exists in harnessing perception for visualization optimization~\cite{micallef2017towards,wang2017perception}, MFMs allow us to tap into a much broader collective knowledge on how humans perceive, and reason about, data visualizations.

Yet an open question remains: how can we best utilize these models for visualization design?
We believe that two distinct sets of problems need to be addressed in order to answer this question.
First, we must question the capability of these models in judging visualizations.
MFMs are trained on \emph{much more} than data visualizations, so it should not be assumed that these models will serve as effective judges.
Thus, we require methods for measuring how well-aligned machine perception is with human perception.
Namely, do machines recognize the same types of visual patterns as humans when observing a visualization?
A better understanding of model capabilities can help set expectations for using MFMs as visualization judges, and help identify what, if anything, we need to adapt models.
Secondly, for the visualization research community to make progress on using MFMs, we need a characterization of what these models offer for visualization design.
A detailed characterization is necessarily model-dependent, thus we consider two types of models: text-to-image (T2I) generative models~\cite{rombach2022high}, as well as multimodal large language models (MLLM)~\cite{liu2023llava}.
We illustrate how these models may be used for design, how they complement one another in achieving design goals of different granularity, and the trade-offs that they present in design vs. optimization flexibility.
We hope that this characterization gives a source of inspiration for researchers in how to approach visualization design with MFMs, both in the moment, and in the foreseeable future.



\section{How effective are multimodal foundation models in perceiving visualizations?}
\label{sec:effective}

MFMs are a product of the data on which they were trained.
Specifically, T2I generative models are commonly trained on captioned images~\cite{schuhmann2021laion}, while MLLMs are usually trained on images associated with instruction-based commands~\cite{liu2023llava}\footnote{Closed models, such as GPT4, do not disclose the data on which they are trained and so we cannot make the same type of claims.}.
As these training datasets are usually assembled via crawling the internet, they are primarily composed of images depicting our surroundings, e.g. outdoor scenes, person-made environments, people, animals, etc.
In contrast, data visualization is relatively niche, and thus the relative frequency of visualization images is expected to be quite low.
And since the effectiveness of foundation models can often be attributed to the training data~\cite{bommasani2021opportunities, yin2023survey, mccoy2023embers}, one might think that multimodal foundation models are unsuitable for visualization, and in particular reasoning about visualization.
Thus, we should first ask: what do multimodal foundation models know about visualization?

\paragraph{Text-to-image generative models}
Take the T2I-based diffusion model SDXL~\cite{podell2023sdxl} as an example.
\addtolength{\columnsep}{-15pt}
\begin{wrapfigure}[14]{r}{0.23\textwidth}
\centering
\includegraphics[width=0.23\textwidth]{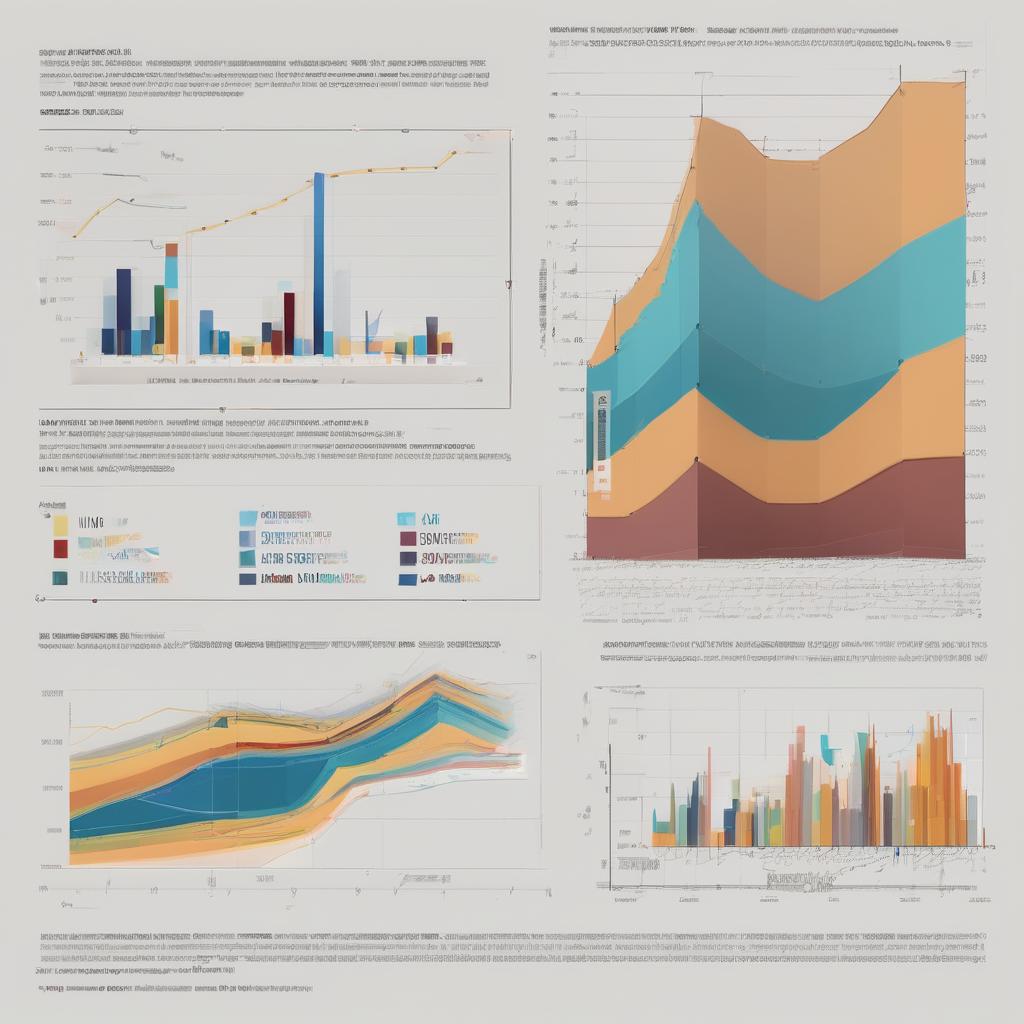}
\end{wrapfigure}
We should expect ``scatterplot'' to be well-recognized by the model, as this is a fairly common type of visualization.
Thus, it should be possible to generate (not particularly useful) images of scatterplots, when prompted with ``scatterplot''.
But what about more elaborate visualizations?
Does a T2I model have a good understanding of, say, multiple-view visualization designs?
In prompting SDXL with ``A clear and concise juxtaposed multiple-view visualization of data'', we obtain the following image shown in the inset.
We observe a multiple-view visualization consisting of stacked bars and stacked area marks, along with consistent color-encoding across views.
Given that we did not specify a dataset, the titles, axes, and legends have no meaning, nor do the length/area encodings.
As we discuss in more detail in Sec.~\ref{sec:t2i}, for the purpose of optimizing visualization parameters, T2I models are better positioned as density models rather than generative models.
Nevertheless, this example indicates that T2I models carry information about nontrivial visualization designs, despite their training data being predominantly natural images.

\paragraph{Multimodal large language models}
MLLMs represent a major recent development in reasoning over vision and language modalities.
In contrast to joint embedding models like CLIP \cite{radford2021learning} that encode image and text in the same latent space, MLLMs can interpret both text and image inputs and more.
However, these models present similar challenges as visualization input likely represents a small fraction of training data. 
Despite these potential challenges, MLLMs can be extremely versatile in their tasks, for example, we can simply ask questions for a given visualization as we see here.
\begin{wrapfigure}[14]{r}{0.28\textwidth}
\centering
\vspace{-2mm}
\includegraphics[width=0.28\textwidth]
{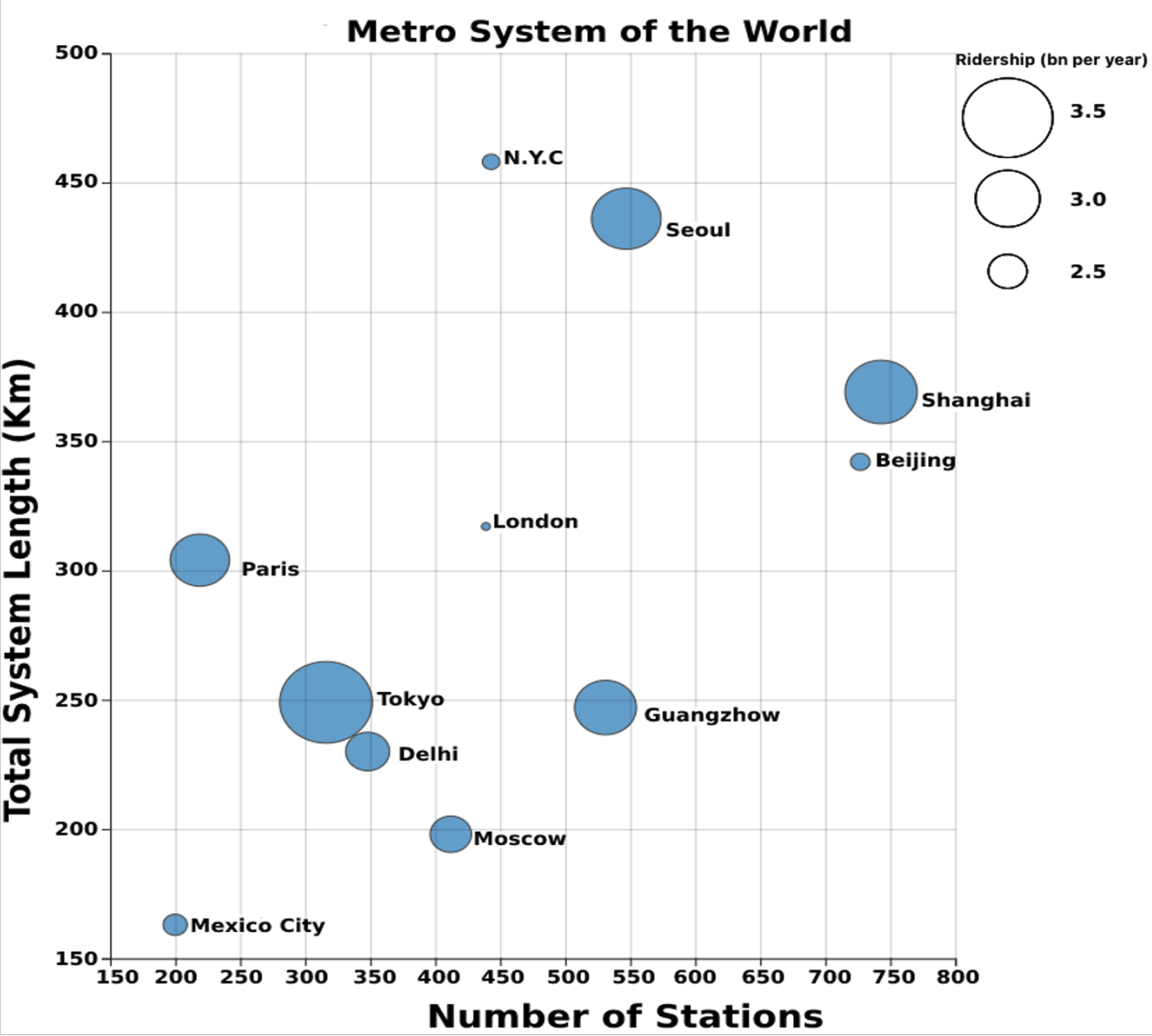}
\end{wrapfigure}
``Based on the given visualization, Which city’s metro system has the largest number of stations?'', this is a test question from Mini-VLAT \cite{pandey2023mini} for testing human visualization literacy. 
The following answer is provided by the model:
\emph{``Based on the visualization, Shanghai has the most metro stations. This can be inferred from its position on the x-axis, which represents the number of stations, where it is the farthest to the right compared to other cities.''}
The MLLM can interpret the image understand the question, and formulate an answer in a human-like way.

\subsection{Measuring human-AI alignment in perceiving visualizations}
\label{sec:human_ai_alignment}

The preceding examples suggest that MFMs encode knowledge about data visualization, and can potentially reason about visualizations.
However, more comprehensive studies need to be conducted to fully characterize the capabilities of MFMs for visualization reasoning.
Indeed, numerous benchmarks~\cite{yang2023dawn, yang2023dawn,masry2022chartqa, mathew2022infographicvqa, xia2024chartx, wei2024mchartqa} exist that can be used to ``test'' MFMs in simple visualization tasks.
These tasks include: determining the value encoded by a graphical mark, value comparisons, trend assessment, and finding extrema.
Moreover, the tasks span different types of graphical marks (point, bar, line), as well as visual channels (position-based encoding vs. length-based encoding).

Although these studies are useful for testing MFMs about visualizations, if we wish to test how well an MFM can perceive visualizations, then more work needs to be done.
Below, we list some potential research avenues.
\begin{enumerate}[wide, labelindent=10pt]
\item \textbf{Summary extraction:} given a large set of marks encoding data items of a quantitative attribute, numerous studies~\cite{szafir2016four,yuan2019perceptual,kale2020visual} have shown that humans are capable of efficiently extracting summaries (e.g. distributional properties such as mean and variance).
``Efficiency'' here means that a human does not need to serially scan each mark in the display, but rather, they process marks in parallel, and can make a judgment in a short amount of time.
How effective are MFMs in extracting summaries?
Prior work~\cite{xia2024chartx,wei2024mchartqa} considers only simple charts for deriving averages.
Thus, we argue that more extensive studies should be done for testing MFMs, for instance varying graphical marks, visual channels, and the number of data items, amongst other factors.
Importantly, \textbf{control tasks} need to be developed to ensure that a model is, indeed, performing ensemble processing, and \emph{not} taking shortcuts.
For instance, the mean could be derived by extracting the value of each mark, summing the values, and dividing by the number of marks.
A control task such as ``count the number of marks'', and the model's ability to complete the task, could then tell us whether the model is processing marks as a whole, rather than individually.
\item \textbf{Preattentive features:} the ability of humans to rapidly identify visual patterns, at-a-glance, is a major impetus for visualization~\cite{healey2011attention}.
This type of bottom-up attention -- pattern detection, independent of a directed task -- is what commonly leads us to ask higher-level, analytical questions about our data~\cite{ware2019information}.
Thus, we believe it is important to measure how well an MFM can preattentively identify visual features, and specifically, \emph{what} visual features are preattentive.
Of course, the more traditional test for humans -- identifying a pattern in approximately 200ms -- does not extend to machines.
To test for this, how to prompt the model becomes essential.
General questions about salient visual patterns can help elicit what humans would rapidly perceive as visually-distinct features.
\item \textbf{Just-noticeable difference:} it is well-studied~\cite{cleveland1984graphical,szafir2017modeling} that humans are limited in accurately noting differences in certain types of visual features, e.g. whether or not two colors are the same, two bars are of the same length.
This further extends to ensemble processing, for instance, whether two scatterplots depict the same level of correlation~\cite{harrison2014ranking}.
We believe that similar notions of just-noticeable difference (JND) should be developed for MFMs.
Given that these models are provided raw pixel values as input, it might seem that JND, at least for simple visual features (e.g. color), does not make sense.
However, in practice, these models use vision image transformers~\cite{dosovitskiy2020image} (ViT) to extract visual representations, giving high-dimensional feature vectors arranged on a coarse spatial grid.
Thus, low-level visual queries might remain a challenge, so methods for measuring JND in MFMs, e.g. analogs to staircasing in human studies~\cite{rensink2010perception,harrison2014ranking}, remain applicable.
\end{enumerate}

\subsection{Adapting model capabilities}
As discussed previously, one of the central challenges of leveraging MFMs for visualization tasks is the lack of significant visualization data in their training distribution, which can lead to limited capabilities.
On the other hand, as discussed in Section \ref{sec:human_ai_alignment}, even a model with advanced vision capabilities is unlikely to be a reliable surrogate for human users due to their potentially disparaging behaviors.
Moreover, even if the vision capabilities of a model are well-aligned to human perception, an incompatibility might exist between \emph{how} visualization goals/tasks are specified by humans, and understood by models.
These challenges, e.g. limited vision capability on visualization and human-AI misalignment, can be mitigated by adapting the model's capabilities to the desirable target use case. 

The most straightforward solution is utilizing specially curated data for task-specific fine-tuning. For example, several works have explored utilizing MLLMs for aesthetic judgment \cite{huang2024aesbench, huang2024aesexpert, hullman2023artificial}. The model achieves such capability through training on data and the corresponding annotation from human users, essentially trying to mimic human assessment, therefore, aligning their preferences.
Similarly, Hao et al. \cite{hao2024visltr} fine-tune the CLIP model~\cite{radford2021learning} for aligning text description, sketch, and chart visualization, for trend querying tasks in tabular data.

However, fine-tuning is not without its limitations, such as catastrophic forgetting, which can be particularly problematic for more capable models as they may significantly harm other capabilities of the original model.
Besides the inherent limitation of fine-tuning, as the model capability grows, it is not unreasonable to expect the model to match or even surpass the human visual capability in certain regards. In such scenarios, creating a persona to better mimic human behavior through prompting, or providing examples that demonstrate the preferred behavior through in-context learning, can help direct the model to the intended capabilities.

\section{Multimodal Foundation Models as Visualization Judges}

Once we know the limits of MFMs in reasoning about visualizations, and (optionally) adapting them to carry certain capabilities, we can be more informed about where these models would be useful in guiding the visualization design process.
But how \emph{should} these models be used as guides?
We do not think it is wise to completely replace visualization designers with MFMs; there should always be a place for human creativity.
Rather, designers should have the flexibility in specifying what they would like an MFM to perform.
To make such a specification natural, an MFM should act in much the same way that a visualization designer acts.
Specifically, \emph{critiquing visualizations} is fundamental to visualization design.
Identifying the limitations of design choices informs what actions we subsequently take.
An MFM can behave in much the same way, wherein we view an MFM as a \textbf{visualization judge}.
Adopting a court analogy, the judge is given information about a case (\emph{data}, \emph{goals of a user}), and processes the evidence (\emph{the visualization}).
In turn, a judgment is made (\emph{an analysis of the visualization, in relation to the goals of the user}).
Last, the judge makes a ruling (\emph{the actions to take in addressing the user's goals}).
In the remainder of this section, we make the above notions more concrete, in the context of T2I-based models and MLLMs.
But first, we more precisely define the models under study.

\paragraph{Definitions}
A \textbf{T2I} generative model accepts as input a text description, and outputs an image that is intended to be compatible with the description.
Mathematically, we can frame this as drawing an image $I$ from a conditional probability distribution over images $I \sim p(I | c)$, conditioning on the text description denoted $c$.
This probabilistic perspective will be useful in purposing T2I models as density models.
A \textbf{MLLM} accepts as input both an image and text and produces text as output.
These are typically autoregressive models, such that given an image and text, they: (1) form a distribution over a discrete vocabulary, (2) draw a sample (i.e. a word) from the distribution, and (3) append the sampled output to the input text.

\subsection{Input (evidence)}
\label{subsec:input}

The input to an MFM is (1) an image of a visualization, and (2) a text description.
We assume that the image is produced based on a dataset, and a visualization specification.
We note that an image is just one way to represent the visualization design itself; the image \emph{can} be equivalent to the dataset \& specification, but might also contain \emph{less information}, especially if the drawn marks are occluded by one another.
Yet, the amount of information within a representation matters little if the model cannot meaningfully reason about the given representation. 
For example, the recognition of clusters in a 2D scatterplot of $2,000$ data points is much easier for humans to perform, compared to inspecting the raw $2,000$ data items, e.g. presented as tabular data.
Although LLMs can, in principle, reason over long contexts, they may struggle to effectively utilize the given information~\cite{li2024long}.
Thus, assuming that MFMs can adequately reason about visualizations (c.f. Sec.~\ref{sec:human_ai_alignment}), then an image-based representation stands to be a more effective representation than the raw data \& visualization specification.

\begin{figure*}[!t]
    \centering
    \includegraphics[width=.94\linewidth]{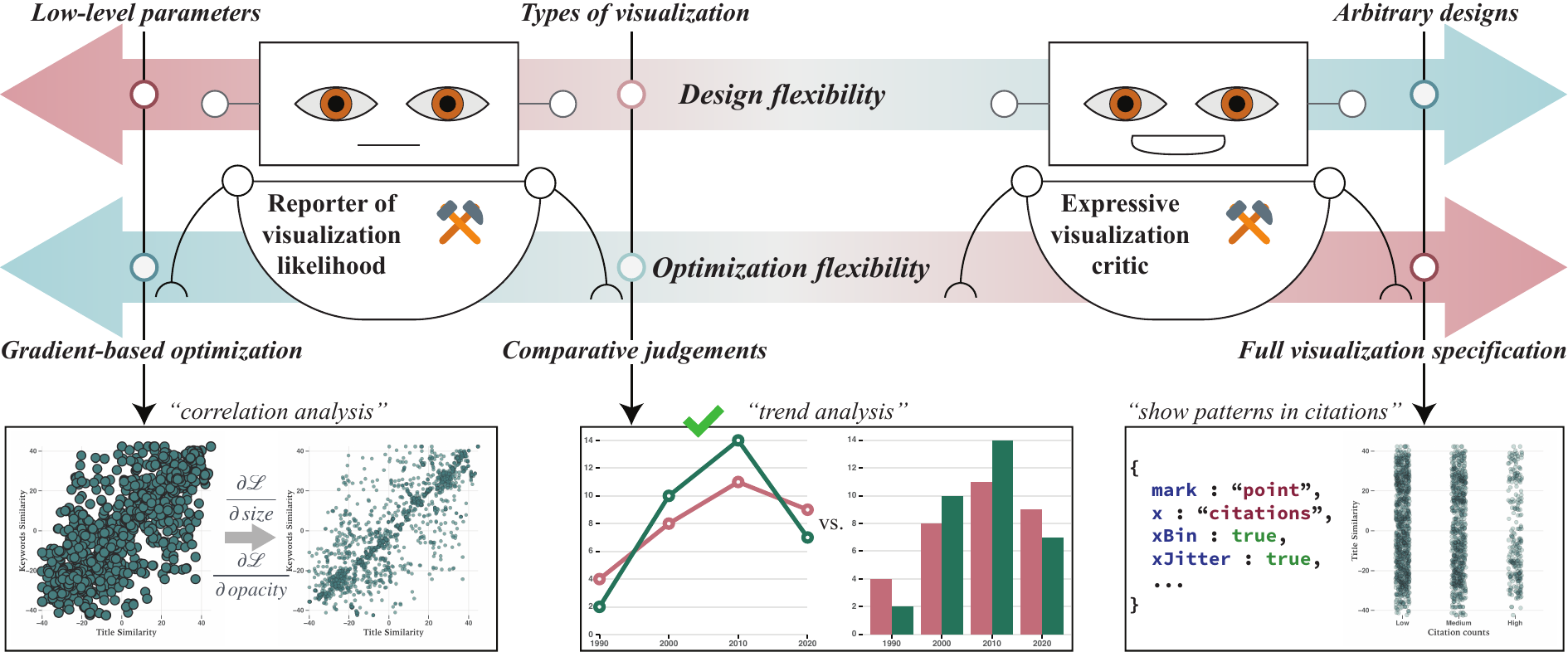}
    \caption{The range of MFMs for visualization design presents a trade off between (1) constraints on the design space, and (2) methods available for optimization. For instance, T2I models that report the likelihood of a visualization offer high flexibility in optimization (gradient-based methods), but are limited in what can be optimized (low-level parameters such as mark size \& opacity). Conversely, as MLLMs are more expressive in their outputs, they permit exploring much more of the visualization design space, but our choices for optimization within this space are more limited.}
    \label{fig:design-optimization}
\end{figure*}

The text supplied to the model should be an encapsulation of the user's goals.
Yet, there are different levels of abstraction in the notion of a ``goal''.
This is related to goal/task abstraction \& taxonomies~\cite{brehmer2013multi,meyer2015nested}, and ways to structure a description of a visualization~\cite{lundgard2021accessible}.
These factors are important, since the text that should be supplied is model-dependent: the input should be based on the model's functionality, e.g. what it is trained to do.
T2I generative models are trained on captioned images, where captions describe the content of an image.
As such, the text input to T2I should describe an object, structure, or more broadly visual patterns that we expect to see in the image -- this constitutes a low-level goal abstraction, a specification of recognizable patterns that can be related with higher-level analyses.
In the case of MLLMs, the input text could be much more broad, referencing not only visual patterns, but attributes in the data, and higher-level analytical goals.
Unlike T2I models, MLLMs contain more general knowledge and have the ability to retain and reason about the context, and history, of interactions.
This allows these models to follow a set of pre-determined steps based on user instruction, and evaluate the goal at each step iteratively.


\subsection{Analysis (judgment)}

Once we supply an image and text to an MFM, it will next perform an analysis of its input.
The output of this analysis depends on the specifics of the model.
As mentioned previously, a T2I model in its more traditional form generates images, given only text.
But to utilize this model as a visualization judge, we are really more concerned with the matter of density estimation, rather than generation.
Not all generative models report density, e.g. generative adversarial networks~\cite{goodfellow2020generative, karras2019style}.
However, methods such as normalizing flows~\cite{papamakarios2021normalizing} or diffusion-based models~\cite{rombach2022high, batzolis2021conditional} give us a way to quantify \emph{how likely} a given image is, either exactly (normalizing flow) or approximately (diffusion model, via a variational lower bound).
Importantly, T2I models yield conditional distributions~\cite{batzolis2021conditional}, conditioned on a text description, and so we can customize the density estimate to the user's goals.
Thus for T2I, given an image and text, the model provides us a continuous score indicating how well the image is described by the text.

For MLLMs, how the input is phrased will influence the structure of the output.
Similar to a T2I model, we may ask for a continuous score, e.g. between $[0, 1]$, indicating how likely a user's goals are satisfied, though unlike T2I, the scores cannot necessarily be treated as well-calibrated probabilities.
Alternatively, we may ask for a binary response, or provide multiple options for the model to choose.
It is also possible to let the model give a free-form text response.
However, it is important that the output remains \emph{actionable}, namely we have the ability to update our design choices, as we discuss next.

\subsection{Action (ruling)}

The output of an MFM gives us the opportunity to update our visualization design.
Going back to the court analogy, if the visualization judge finds us ``guilty'' (a bad visualization), then its ruling will, ideally, give us a way to correct for our misdeeds (make our visualization better).
How this is operationalized in an MFM is model-dependent.
For a T2I model, let's suppose it provides us a continuous score, in the form of a log likelihood $\log p_{\theta}(I | c)$, where $\theta$ represents the model's parameters.
As deep networks are (usually) differentiable functions, then this implies we can compute gradients of the log likelihood $\nabla \log p_{\theta}(I | c)$ -- \emph{acting} on the gradients can bring us closer to the user's objectives, giving us a higher likelihood.

MLLMs do not offer such detailed information as gradients, since they output text, rather than a numeric score.
Yet, just as language models are autoregressive, we need a similar mechanism for visualization, e.g. we can make an update to our visualization, given the output of the MLLM.
This presumes a representation of the visualization design space has been created, by the visualization designer, giving (1) a space of possible actions to take, and (2) a means of selecting an action given the model's output.

\subsection{Trade-offs in design and search}

Equipped with a means of judging visualizations, and taking action, an MFM can search the visualization design space in the pursuit of finding effective visualizations.
However, dependent on the specific model, trade-offs exist between the size of the visualization design space, and the effectiveness of the methods available for searching (namely: optimization methods).
Fig.~\ref{fig:design-optimization} depicts an illustration of these trade-offs.
By ``design space'', we mean all possible options for visually encoding data, ranging from low-level parameters that govern the appearance of graphical marks (Fig.~\ref{fig:design-optimization}-left), to choices on what should be shown, e.g. data transformations (Fig.~\ref{fig:design-optimization}-right).
In what follows we characterize these extremes.

In considering the output of an MFM, high-level design choices can be easily expressed through an MLLM, as text is produced as output.
However this comes at a cost: given the current visualization, optimization is limited by the model \emph{directly predicting} what the next visualization should be.
Ideally, the model can utilize the current visualization to make an \emph{improvement} on the user's goals.
But within this process, the (human) designer is removed from optimization.
We are reliant on the model's text output as the main signal in finding the best solution.

In contrast, if we require much less from the model, e.g. a score of how likely a visualization is given the user's goals, then it is not apparent what actions we should take.
However, let's suppose we limit our design space, for instance to choices on how we encode data attributes with visual channels.
Then we can utilize this likelihood signal to give us a \emph{direction} on where to move in this space, in order to improve the visualization, e.g. the parameters of the encoding map.
Thus search amounts to coupling the model's output with optimization procedures created by the (human) designer, for instance gradient-based optimization.

Beyond these extremes, considering what is possible throughout the whole spectrum, we believe, can inspire more diverse research directions.
For instance, as shown in the center of Fig.~\ref{fig:design-optimization}, comparative judgements of visualizations can give a different type of signal, and lead to different flavors of optimization algorithms.
Nevertheless, at the moment, we have characterized MFM-guided visualization design at a rather high level.
Thus in the proceeding sections, we give concrete examples and more deeply characterize how MFMs can be put to use for designing visualizations.

\section{Visualization Design with Text-to-Image Generative Models}
\label{sec:t2i}

As previously mentioned, T2I generative models, purposed as density estimators, have the potential to be used as part of the design process.
But in practice, what needs to be done to realize this potential?
In this section, we outline the general problem of \emph{optimizing visualizations} using text-based maximal likelihood.

\subsection{Problem Formulation}

We first provide an abstract mathematical formulation for the problem.
Suppose that we are given a dataset, denoted $\mathcal{D}$, that we wish to visualize.
We assume that the user has settled on a particular space of visualization designs, and each design can be identified by parameters, represented as a real-valued vector $\mathbf{v} \in \mathbb{R}^D$ (we will loosen this assumption to a mixture of constrained, or discrete, variables in Sec.~\ref{subsec:diff-vis}).
We represent this process of visualizing data as a function $V$, one that accepts two inputs, (1) the dataset $\mathcal{D}$, and (2) visualization parameters $\mathbf{v} \in \mathbb{R}^D$, and outputs an image $I = V(\mathcal{D}, \mathbf{v})$.
Last, we assume that a user has specified a goal for the visualization, represented in the form of a text description denoted $c$.

The goal of optimization is to find parameters $\mathbf{v}$ that lead to a visualization $I = V(\mathcal{D}, \mathbf{v})$ which is compatible with the user's description $c$.
We can measure the ``compatability'' between a visualization and a text description through a model that can report \emph{how likely} a given image is, conditioned on text.
Certain generative models can not only synthesize images, but also report exact~\cite{papamakarios2021normalizing} or approximate~\cite{li2023your}, likelihoods.
We can purpose such models for optimization, and ask for visualization parameters whose visualization images maximize the (log) likelihood:
\begin{equation}
\argmax_{\mathbf{v}} \log p_{\theta}\left(V(\mathcal{D}, \mathbf{v}) \, | \, c \right),
\label{eq:maxlik}
\end{equation}
for model $p$ that reports the likelihood of an input image, with model parameters $\theta$.

This represents a less conventional approach in how to use generative models.
Indeed, for visualization design, their generative capabilities have been used for stylizing visualizations~\cite{schetinger2023doom,wu2023viz2viz,zhao2024leva}.
But our proposed formulation leads to an inverse problem: we are using a generative model as a \emph{prior}, finding a visualization image that maps into the model's data distribution.
An important distinction, however, is that we are not finding just \emph{any} image, but rather, we are restricting the space of visualizations by the parameters, $\mathbf{v}$, on which we are optimizing.
We take inspiration from recent works in using diffusion models as priors~\cite{chung2023solving,mardani2023variational}, for instance, finding radiance fields of 3D scenes solely from a text description specifying the scene~\cite{poole2023dreamfusion,epstein2024disentangled3dscenegeneration}.
Yet an important distinction is that, for visualization, we can (and should!) frame optimization in a much more constrained manner.
Specifically, we should limit optimization to a particular space of visualization parameters, while fixing the rest of the visualization to decisions made by the designer.

\textbf{User-specified goals of a visualization.} Another important distinction from 3D-based inverse problems lies in the specified text description.
There is a big difference in how one would describe a natural scene, and how one would describe the objectives that their visualization should satisfy.
To bring the most benefits for visualization design, we argue that text descriptions should be representative of \emph{hypotheses} that end users have about their data.
Building on Sec.~\ref{subsec:input}, the user might have a hunch about visual patterns that might be present, for instance correlation, clusters, increasing/decreasing trends, spatial density, outliers, etc.
These patterns can then be expressed via a text description, and by solving Eq.~\ref{eq:maxlik}, we may then find parameters whose visualization images satisfy the user's communicated hypotheses.

\subsection{Examples}

In order to make the general problem formulation more concrete, we list a series of examples of exactly what aspects of design we may optimize.
Common to all examples is the definition of visualization parameters $\mathbf{v}$; once we have defined these parameters, and a means of producing an image $I = V(\mathcal{D}, \mathbf{v})$ from the dataset and parameters, then we can perform optimization (c.f. Eq.~\ref{eq:maxlik}, see also Sec.~\ref{subsec:diff-vis}).

\textbf{Data-independent design choices.} When a designer is making choices about a visualization, arguably the most important choices are concerned with visual encodings of data.
Yet, certain choices are data-independent, which can nevertheless have a significant impact on the resulting visualization.
For instance, if we are encoding data items by individual graphical marks, then the mark size, mark color, transparency of marks, and mark outline are all necessary choices on the visual appearance of marks, even if these visual channels do not encode aspects of the data.
Each of these visual properties can be represented by an individual dimension within our visualization parameter vector $\mathbf{v} \in \mathbb{R}^D$, where $D$ indicates the number of properties.
As such, they can be optimized for a variety of user-specified objectives.

This perspective on parameter optimization is not limited to simple visual properties.
For instance, we can optimize for the \emph{shape} of a graphical mark, whereby shape parameters become our visualization parameters.
This can be a considerably more flexible means of glyph design, in contrast with manual shape editing of glyphs~\cite{ren2018charticulator,liu2018data}.
As an example, if one aims to craft a more memorable visualization~\cite{borkin2013makes}, where graphical marks take the form of recognizable shapes, then expressing these shapes through language becomes straightforward.

\textbf{Visual encodings of data.} To identify the parameters involved in the visual encoding of data, we define an encoding as a \emph{map}: mapping the value of a data attribute to the range of a visual channel.
From this perspective, visualization parameters can be distinguished by the (1) data domain, (2) map properties, and (3) the visual range.
For continuous data, typically the data domain is identified by the minimum and maximum data value.
Though it is convention to set this to the data's extent, optimizing these parameters can be preferable if only certain portions of the data domain present user-specified visual patterns.
The mapping, itself, can further be represented as a function with parameters to be optimized.
For instance, this can be a polynomial function whose coefficients become our parameters, thus giving us a linear mapping at one extreme, and a high-order polynomial on the other.
Which one is more preferable is dependent on the data, and description specified by the user.

The visual range of a mapping offers a number of opportunities for optimization.
In its simplest setting, if the visual range is continuous, then we might only need to decide on an interval, namely a minimum and maximum value.
For instance, were we to use size as a visual channel, then we need only find the minimum and maximum size.
However, color as a (continuous) visual channel deserves special attention.
We can treat the color range as its own parameterized function, defined in such a way to respect basic design principles (e.g. luminance-increasing~\cite{liu2018somewhere}, multi-hue~\cite{reda2020rainbows}), while being optimized to adhere to a user's description.

\begin{figure}[!t]
    \centering
    \includegraphics[width=0.48\textwidth]{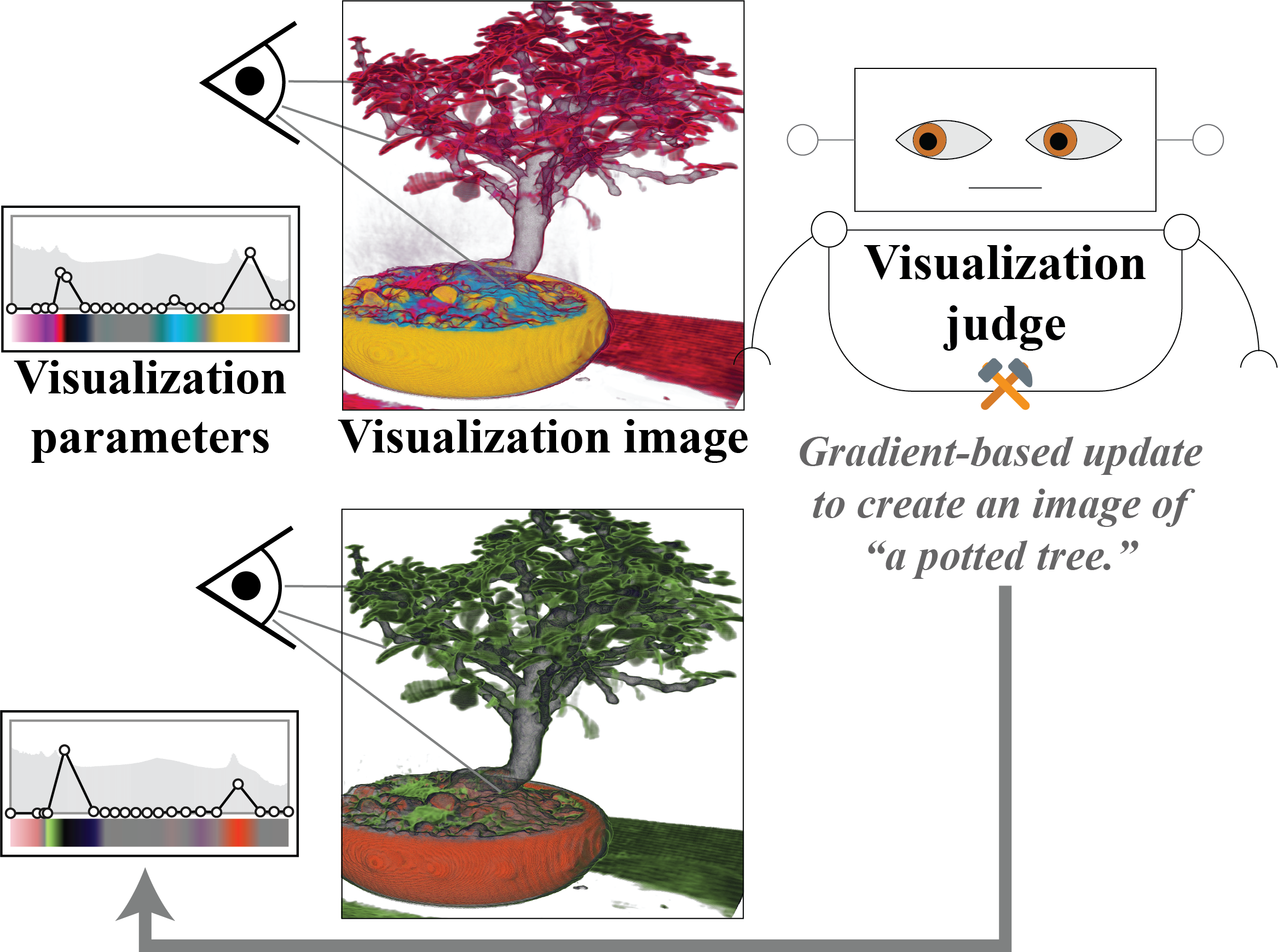}
    \caption{Our prior work~\cite{jeong2024textbased} considers how to find visualization parameters -- transfer functions for color and opacity -- that give visualizations which are compatible with a user's description. Here we show how different styles of this volumetric data, representing a tree, can be depicted through simply changing the descriptions of the volume, or what the user aims to see in the data.}
    \label{fig:text-to-tf}
\end{figure}

As a more concrete example, we show from our prior work~\cite{jeong2024textbased} how color, as well as opacity, can be optimized to satisfy a user's description within the context of volume visualization, please see Fig.~\ref{fig:text-to-tf} for an illustration.
Given volumetric data as input, a common way to control the visualization of the volume is through specifying transfer functions for both color and opacity.
The manual specification of transfer functions, however, can be tedious.
Thus a language-based specification, describing what one would like to see in the volume (e.g. different styles of a tree), is a more intuitive way to explore structures in the data.
Given the text description, the transfer functions can then be found via optimization\footnote{Though our work considers CLIP-based scoring~\cite{radford2021learning} rather than maximum likelihood (c.f. Eq.~\ref{eq:maxlik}), the general principle remains the same.}.

\textbf{View arrangement.} Multiple-view visualization design involves numerous factors, perhaps most importantly, what to show in each view and how to show it.
But another important aspect is the spatial arrangement of views, namely where individual views are positioned, and the size of each view~\cite{chen2020composition}.
Effective utilization of a small amount of drawing space can be a challenge, and thus the ability to optimize for view arrangements can alleviate this burden for the user.
Specifically, view parameters (position, width/height), subject to constraints on the global spatial layout, can be optimized to clearly convey information in each view.
This can have the effect of adapting each view's size to the content of the data, ensuring the user's objectives are satisfied amongst all views.

\textbf{Visualization technique parameters.} Data encodings derived from visualization techniques, e.g. dimensionality reduction (DR)~\cite{van2008visualizing,mcinnes2018umap} or graph layouts~\cite{wang2017revisiting}, often depend on parameters that can significantly impact the output results.
The setting of these parameters, however, can lead to unpredictable results, for instance setting of neighborhood size in DR methods~\cite{van2008visualizing,mcinnes2018umap}.
To this end, the ability to express certain qualities in the output, e.g. clusters in DR, local symmetries in graph layouts, can be used to determine the right set of parameters for visualization techniques.

\subsection{Differentiable Visualization}
\label{subsec:diff-vis}

Thus far we have defined the optimization problem, and specific examples of what we can optimize.
But we have not discussed \emph{how} to perform optimization.
Ideally, we can use gradient-based optimization to maximize Eq.~\ref{eq:maxlik}.
But this requires (1) the likelihood $p_{\theta}$ to be a differentiable function with respect to the image, and (2) the visualization function $V$ to be differentiable, with respect to visualization parameters.
As $p_{\theta}$ often takes the form of a deep network, the likelihood admits derivatives.
Efficiently computing these derivatives, especially for modern T2I models (e.g. diffusion models~\cite{podell2023sdxl}), presents some challenges, but methods such as score distillation sampling~\cite{poole2023dreamfusion} give us practical approximations.
Moreover, if our parameters are not real-valued scalars, then we need to consider constrained optimization.
Specifically, if our parameters are continuous, then mappings that enforce predefined constraints can be used, while for discrete data we can use the Gumbel-Softmax technique~\cite{jang2017categorical} for approximate gradients.
For the visualization designer, then, what remains is the declaration of the visualization function.
We need to ask: how can we optimize over visualizations?

In this setting, \textbf{differentiable visualization} is necessary.
Specifically, every pixel in the output image needs to be a differentiable function of the visualization parameters.
This requires us to think more deeply about the visualization code we write, and how we use the resulting image output.
We need to \emph{rasterize} the graphical marks, axes, labels, and all other aspects of what is shown in a visualization, as an image of fixed resolution that is then supplied to the T2I model.
Just considering graphical marks, usually these are specified as analytically-defined shapes, such as circles, squares, etc.
A raster-based representation that is differentiable wrt our parameters is not easy to derive, considering that as a function over the image plane, the mark's shape induces a step function.
Hence, differentiable proxies become necessary to consider, and we can look toward the graphics community for 2D/3D-based rasterization for inspiration~\cite{liu2019soft,li2020differentiable}.
Here we think some lessons can be learned from existing optimization-based rendering APIs, such as Nerfstudio~\cite{nerfstudio}.




\section{Visualization Design with Multimodal Large Language Models}
In the previous section, we explored utilizing T2I models as density estimators for optimization, and one of its key advantages is providing an explicit gradient signal when combined with a fully differentiable visualization pipeline. 
However, obtaining meaningful gradient signals from MLLM models represents significant challenges due to their autoregressive nature and the output modality. 
On the other hand, due to their inherent knowledge, particularly regarding the ability to retain and reason about history and context, the strength of MLLM lies in their potential to provide overall evaluation and feedback similar to human evaluators. With these potential strengths and limitations in mind, in this section, we explore the extent to which MLLMs can be used for guiding visualization design.

\subsection{Explicit Optimization with MLLM}
The first question we want to answer is whether explicit optimization with MLLMs is possible for continuous visualization parameters, even if we do not have gradient information.
Despite the challenges, such optimization is possible by utilizing MLLM as the judge or rank of the visualization inputs according to the user-stated goal. 
To produce meaningful updates to the visualization parameters, we can leverage zeroth-order optimization \cite{liu2020primer} (e.g., in this case, comparative zeroth-order optimization) to reveal meaning directions for parameter updates. Such a usage scenario can be a powerful way to infuse human preference by combining the expressiveness of natural language and the nuanced knowledge an LLM possesses.

\begin{figure}
    \centering
    \includegraphics[width=0.47\textwidth]{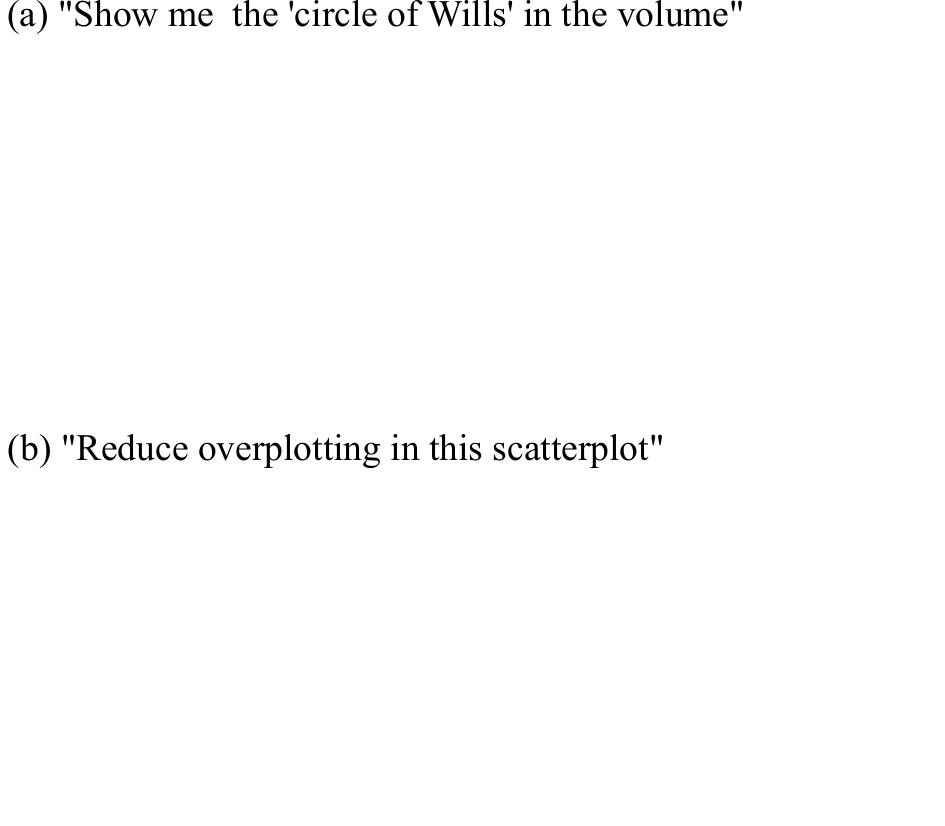}
    \caption{Can an MLLM make adjustments to the visualization parameters automatically based on (1) their visual understanding of the visualization output, and (2) the user instructions in natural language? The results here are the AVA work \cite{liu2024ava}, where we demonstrate an autonomous visualization agent can be designed by allowing the MLLM to iteratively refine the existing visualization.}
    \label{fig:ava_example}
\end{figure}

For example, as illustrated in Figure \ref{fig:ava_example} (from the early work demonstrated by Liu et al \cite{liu2024ava}), the MLLM allows us to link explicit domain-related knowledge with their visual representation which can be hard to achieve otherwise. 
In (a), the model will need to understand what the ``circle of Willis'' is (one of the key arterial structures in the brain \cite{alpers1959anatomical}) and be able to assess whether the rendered image depicts such a structure. 
In (b), the model will need to understand what overplotting means, and what criteria to look for when judging whether overplotting occurs. All these criteria would be almost impossible to be mathematically defined. 

However, such an optimization strategy is not without its limitations. As discussed in the AVA work \cite{liu2024ava}, we expect such optimization to become much harder as the solution space, i.e., the parameter space, increases. One potential mitigation is to narrow down the search space by leveraging domain knowledge similar to how human users fine-tune a given visualization. By understanding the potential impact of the parameters and even building a mental model on the expected outcome, human users can obtain desirable visualizations from a complex multiparameter setting without the need for explicit gradient.

\subsection{Evaluation and Recommendation with MLLM}
In the previous section, we elucidate that even though explicit optimization is possible with MLLM, there are many associated challenges, T2I-based likelihood estimation may provide a much better way for parameter optimization.
The advantage of LLM, and to the extent of MLLMs is the language capability.
Compared to explicit optimization, MLLMs are much better at providing general assessments and recommendations in natural language.
LLM has long been adopted for NLP research in assessing the quality of natural language prediction, particularly for data selection and filtering \cite{zhu2023judgelm, chiang2023can}. A recent study \cite{chiang2023can} suggests that the LLM evaluator performs similarly to the human experts on textual input. 
So naturally we wonder whether MLLM can play a similar role as a visualization evaluator.
Instead of updating lower-level parameters of a specific design, could MLLMs leverage their knowledge and visual perception capabilities to act as a visualization evaluator, and identify the \emph{best} visualization amongst a set of candidates? 

Many existing works have explored visualization recommendation \cite{zhang2023adavis, li2021kg4vis, hu2019vizml, vartak2017towards, zhao2024leva}.
For example, the recent work LLM4Vis \cite{wang2023llm4vis} utilizes GPT-4 to produce visualization recommendations and the corresponding explanation. 
Similarly, Kim et al. \cite{kim2023good} investigate whether LLM can be used to give advice on a visualization specification and tasks.
However, one important distinction is that none of these works rely on visual perception for their evaluation.
In some sense, the model produces the recommendation or judgment of a given visualization without ever assessing it through visual perception.
All the assessments are done through intermediate means, from visualization description (e.g., visual grammar) to data (extracted features and descriptions, or raw data). 
At least for computer vision, prior work~\cite{chen2024analyzing} has shown that language alone remains inferior to vision combined with language, and we expect this to hold for visualization design.

The absence of visual perception for visualization evaluation is a significant departure from how humans approach similar assessment problems.
Essentially, an implicit assumption has been made, i.e., the LLM can envision the resulting visualization so that correct assessments can be achieved without viewing the actual visualization. 
However, we likely need significantly more evidence to determine whether that assumption is true or even theoretically possible.
Often, the ideal choice of visualization is invariably tied to the data and we have yet to observe the strong capability of LLM to directly comprehend patterns from raw data despite the increasing context length.
To make a fully informed evaluation, the model likely needs to `see'' the combined effect of data \& design specification.
Therefore, we believe not leveraging the actual visualization represents a significant missed opportunity, especially considering the growing capabilities of MLLMs. 
The inclusion of visual perception represents the most straightforward process that best resembles how humans approach such tasks.
There are limited works that explore utilizing vision models to assess visualization output. 
Despite focusing on lower-level information retrieval tasks rather than high-level assessment, the VisLTR work \cite{hao2024visltr} explores the use of CLIP~\cite{radford2021learning} for selecting plots with the specific visualization encoding and data patterns to match the provided language description. 

\subsection{Visualization Generation vs. Visualization Design}
So far, we have discussed how MLLM can be adopted for explicit optimization and how MLLM can be used for visualization evaluation and judgment, each with increasing design complexity. As illustrated in Figure \ref{fig:design-optimization}, as we move up further in the design complexity axis, one may wonder whether MLLM can fully design visualization that meets a given specification.

Upon initial examination, this direction may appear widely studied already. Generating visualization has been the primary focus of many LLM-related visualization work \cite{chen2023beyond, tian2024chartgpt, yang2024matplotagent}.
Generating code is something the LLMs can reliably achieve due to their strong capability in text token generation.
As a result, directly generating visualization code, i.e., visualization grammar, python plotting library, or even Javascript visualization, is central for creating a visualization based on a given specification. And that process only requires a language-only model without any vision capabilities.

However, an important distinction that needs to be made here is that visualization generation is not equal to visualization design.
Visualization generation focuses on the more fundamental mechanistic capability, whereas visualization design implies the inclusion of actual application goals as well as a decision process. 
The design carries a clear intent with expected outcomes.
In addition, many existing visualization generation works rely on users to provide general design guidance, e.g., the type of visual encoding.
This deviates from more traditional design, as the intent is for the system to determine the full visualization specifications without constraints or prior instructions.

To arrive at the true visualization design capability that allows an arbitrary design with full specification, the system needs to break down the overall design process, including the conception, implementation, and refinement stages. It needs to figure out what visual encoding to use, implement the visualization, and more importantly, view the visualization on the given data to understand the strengths and weaknesses of the current design, and then potentially go back to the drawing board and re-imagine what the visualization encoding should be. 
The overall process can be best described as an iterative process while utilizing many individual capabilities we have discussed so far. For example, the MLLM can suggest the initial visualization choice, and generate the code to produce the visualization.
Then, its visual perception capability can be utilized to determine the effectiveness of the actual visualization, and for the specific visualization parameters, whereby another model can be employed for low-level gradient-based optimization. 

Visual perception will be critical in forming such an iterative improvement loop.
It would be significantly more challenging to design a visualization without actually observing the visualization visually, even if text-only models can have certain visual knowledge \cite{sharma2024vision} from their training data, such as text in SVG files. 
Moreover, for such a design process to work, an autonomous visualization agent (the general concept is proposed in \cite{liu2024ava}) is likely needed to act as the brain of the operation, where multiple components are coordinated to work together, each potentially employing different models. 
%



\section{Discussion}

Looking ahead, as MFMs rapidly progress in capabilities, it is difficult to predict how these models could be purposed to support users in visualization design.
Nevertheless we hope that our characterization of MFMs are useful, both in the moment, and future models, for inspiring visualization researchers on utilizing MFMs for design.
We believe that we are only scratching the surface in how to utilize these models, however.
Thus we leave with some concluding thoughts on MFMs for visualization design.

\textbf{Visualization robustness.} Consider the following scenario: given a dataset, we form two separate datasets, each produced via sampling data items from the original, drawn uniformly at random with replacement.
For each dataset, we then use an MFM to find corresponding visualizations, optimized for the same user-specified objective.
However, we find that the resulting visualizations are quite different: one addresses the user's objective, but we perceive the other as a failure case, despite the MFM reporting a good finding.
We anticipate such brittleness to (presumably) small changes in the data to be common.
As a consequence, more robust methods for optimization will need to be developed to handle these problems.
For instance, we should be seeking designs for which an MFM's judgement is \emph{consistent} amongst a set of visualizations, each of which is based on a random sampling of the data (akin to bootstrapping).
Ideally this can mitigate failure modes of the model, and give designs that are robust to small changes in the data.

\textbf{Finding a diverse set of designs.} It might be hubris to approach visualization design as finding a single, optimal visualization.
Visualization is subjective~\cite{van2005value}; a visualization considered ``optimal'' by one person might not be for another person.
Finding the ``best'' visualization therefore might be a poor strategy.
On the other hand, our characterization of visualization judges, and the ability to take actions in a design space, can be utilized for gathering a \emph{set} of visualizations, all of which are judged as good by the model.
We believe this requires a shift in perspective: from gradient-based optimization, to Bayesian inference~\cite{gelman2013bayesian}.
The aim would be to sample from the posterior distribution of visualization parameters, conditioning on data and user objectives.
Similar approaches have been taken for finding layouts in multiple-view designs~\cite{shao2021modeling}; we advocate for inference in using MFMs to sample a design space.

\textbf{Combining text and images to express a user's goals.} In certain circumstances, encapsulating a user's goals via a text description is inconvenient, if not impossible.
Often designers obtain inspiration from existing visualizations~\cite{sedlmair2012design}, used as reference for their own problem setting.
Alternatively, designers will sketch, on paper, how they envision their visualizations should look~\cite{bigelow2014reflections,roberts2015sketching}, e.g. what individual views contain, the spatial arrangement of views, and how views are linked together.
Thus images can play a crucial role in helping the user express what they want in a visualization.
Combining text-based descriptions and image-based descriptions is thus a fruitful research direction in using MFMs to guide design.

\textbf{Visualization design for machine comprehension.} Conventionally, visualizations are consumed by humans in the support of data analysis~\cite{van2005value}.
But what if the end user is no longer a human, but rather, a machine, one that still aims to perform data analysis?
Of course, images of visualizations are not necessary for language models to perform reasoning.
However, providing raw data as input is limited by the model's ability to use long-form context (c.f. Sec.~\ref{subsec:input}).
Moreover, what is considered a good design for a human might not be a good design for a model (c.f. Sec.~\ref{sec:human_ai_alignment}).
Visualization design thus gets flipped on its head: we need to consider optimal visualization designs for machines, rather than humans.
Recent work has in fact shown that encouraging MLLMs to explicitly visualize their thoughts~\cite{wu2024visualization} can improve on their problem solving abilities.
We believe this line of work can be extended towards data analysis, but with humans engineering how machines should design visualizations for their own analytical thinking.

\section*{Acknowledgements}
This work was performed under the auspices of the U.S. Department of Energy by Lawrence Livermore National Laboratory under Contract DE-AC52-07NA27344 and was supported by the LLNL-LDRD 23-ERD-029. Released under LLNL-PROC-866816.
\bibliographystyle{abbrv-doi}

\bibliography{vis}
\end{document}